\documentclass[%
 reprint,
 amsmath,amssymb,
 aps,
floatfix,
]{revtex4-2}

\usepackage{graphicx}% Include figure files
\usepackage{dcolumn}% Align table columns on decimal point
\usepackage{bm}% bold math
\usepackage{xcolor}
\usepackage[acronym]{glossaries}
\usepackage{hyperref}% add hypertext capabilities
\usepackage[caption=false]{subfig}
\newacronym{lisa}{LISA}{Laser Interferometer Space Antenna}
\newacronym{tdi}{TDI}{time-delay interferometry}
\newacronym{tm}{TM}{test mass}
\newacronym{eh}{EH}{electrode housing}
\newacronym{gw}{GW}{gravitational wave}
\newacronym{esa}{ESA}{European Space Agency}
\newacronym{lpf}{LPF}{LISA Pathfinder}
\newacronym{mcmc}{MCMC}{Markov Chain Monte Carlo}
\newacronym{ldc}{LDC}{LISA Data Challenge}

\begin{document}

\newcommand{\todo}[1]{{\color{red}TODO: #1}}
\newcommand{\orcid}[1]{\href{https://orcid.org/#1}{\includegraphics[width=8pt]{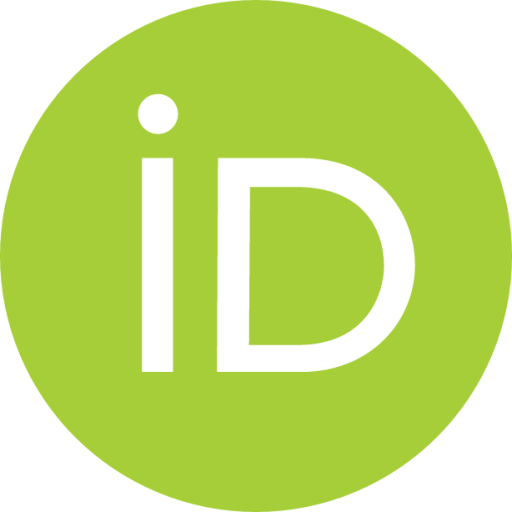}}}
\preprint{APS/123-QED}

\title{Maximum Likelihood Detection of Instrumental Glitches in LISA TDI Data}

\author{Orion Sauter \orcid{0000-0003-2293-1554}}
\email{orionsauter@ufl.edu}
\affiliation{Department of Mechanical and Aerospace Engineering, MAE-A, P.O. Box 116250, University of Florida, Gainesville, Florida 32611, USA}

\author{Peter Wass \orcid{0000-0002-2945-399X}}
\email{pwass@ufl.edu}
\affiliation{Department of Mechanical and Aerospace Engineering, MAE-A, P.O. Box 116250, University of Florida, Gainesville, Florida 32611, USA}

\author{Wiler Sanchez \orcid{0000-0001-9014-6254}}
\email{wilersanchez@ufl.edu}
\affiliation{Department of Mechanical and Aerospace Engineering, MAE-A, P.O. Box 116250, University of Florida, Gainesville, Florida 32611, USA}

\author{Henri Inchauspé \orcid{0000-0002-4664-6451}}
\email{henri.inchauspe@kuleuven.be}
\affiliation{Institute for Theoretical Physics, KU Leuven,
Celestijnenlaan 200D, B-3001 Leuven, Belgium}
\affiliation{Leuven Gravity Institute, KU Leuven,
Celestijnenlaan 200D box 2415, 3001 Leuven, Belgium}
\affiliation{Institut f\"ur Theoretische Physik, Universit\"at Heidelberg, Philosophenweg 16, 69120 Heidelberg, Germany}

\date{\today}% It is always \today, today,
             %  but any date may be explicitly specified

\begin{abstract}
The orbiting LISA instrument is designed to detect gravitational waves in the millihertz band, produced by sources including galactic binaries and extreme mass ratio inspirals, among others. The detector consists of three spacecraft, each carrying a pair of free-falling test masses. A technology-demonstration mission, LISA Pathfinder, was launched in 2015, and observed several sudden changes in test mass acceleration, referred to as ``glitches.'' Similar glitches in the full LISA mission have the potential to contaminate the Time-Delay Interferometry outputs that are the detector's primary data product. In this paper, we describe an optimization technique using maximum likelihood estimation for detecting and removing glitches with a known waveform.
\end{abstract}

%\keywords{Suggested keywords}%Use showkeys class option if keyword
                              %display desired
\maketitle

%\tableofcontents

\section{\label{sec:intro}Introduction}
The \gls{lisa} is a planned space-based gravitational wave detector. It will consist of three spacecraft in heliocentric orbits, exchanging 6 laser links \cite{lisa}. \gls{lisa} is designed to detect waves in the millihertz band using a pair of free-falling \glspl{tm} contained in each spacecraft. \Glspl{gw} in this band are expected to be produced by galactic binaries, massive black hole binaries, and stochastic background, among other sources\cite{LISAReview}.

As a technical demonstration of the \gls{tm} free-fall and drag-free control required for \gls{lisa}, the \gls{esa} launched \gls{lpf} in 2015. The mission showed better than expected free fall performance \cite{LPFResults}, but also included a number of ``glitches'' in the \gls{tm} acceleration. The majority of these were determined to result from residual gas trapped in the device releasing suddenly \cite{GlitchSource}. Similar glitches are likely to occur in \gls{lisa}, and so we wish to examine their potential effects on the \gls{tdi} data products.

Previous papers have discussed more general techniques for detecting glitches. Baghi et al. used a shapelet model with matched filtering \cite{GlitchDetect} in successive steps to reduce the glitch to the level of random noise. That technique was used for both the glitch waveform examined here, as well as sine-Gaussian burst signals, which we aim to avoid flagging. Houba et al. modeled generalized glitches using a machine-learning approach \cite{GlitchML}. However, that paper notes the need for retraining the model for different injection locations and parameters. By focusing on a specific glitch model with a known source, we can simplify the detection procedure, while still improving the data to a glitch-free state.

Due to the density of signals in the \gls{lisa} band, we must use a global fit to extract the various sources \cite{globalfit}. The fit is performed with a \gls{mcmc} routine that varies the number and types of signals present in the data. Since many signals overlap, in particular the galactic foreground, all sources must be considered simultaneously. The glitch model presented here can easily be adapted to the existing global fit as another signal type, with parameters that can be added to the full set.

We describe the model used for glitches in Section \ref{sec:glitch}, and their resulting \gls{tdi} response in Section \ref{sec:tdi}. In Section \ref{sec:likelihood} we describe the maximum likelihood technique used to detect glitches in \gls{tdi}, and Section \ref{sec:gendata} outlines how those \gls{tdi} timeseries were generated. Finally, in Section \ref{sec:results} we give the results of the search, and conclusions in Section \ref{sec:concl}.

\section{\label{sec:glitch}Glitch Model}
\begin{figure}
    \centering
    \includegraphics{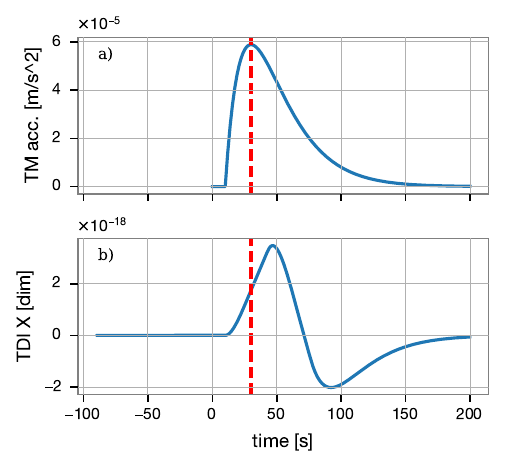}
    \caption{(a) \gls{tm} acceleration for a glitch with $\beta = 20$ s, $\Delta v = 2\times 10^{-8}$ m/s, and $t_0 = 10$ s. Dashed line shows time of maximum acceleration, $t = t_0+\beta$. (b) TDI X response to the same glitch.}
    \label{fig:gaccel}
    \label{fig:glitchtdi}
\end{figure}

The impulse-carrying glitches described in \cite{GlitchSource} have an acceleration profile given by
\begin{equation}
    \label{eqn:gaccel}
    \ddot{x}(t) = \frac{\Delta v}{\beta^2}t e^{-t/\beta} \Theta(t),
\end{equation}
where $\Delta v$ is the total change in velocity imparted to the \gls{tm}, $\beta$ is the time of the maximum acceleration (i.e. $\ddot{x}(\beta)=\max\ddot{x}$), $\Theta$ is the Heaviside step function, and $t$ is the time since the start of the glitch. The function's shape consists of a steep rise up to a maximum value, followed by a decay back to zero, with duration determined by $\beta$. An example is shown in Figure \ref{fig:gaccel}.

This model is explained in \cite{GlitchSource} as a release of gas molecules from a specific point in the \gls{eh} that surrounds the \gls{tm}, resulting in a steep rise in acceleration, followed by the molecules diffusing in the space between the two surfaces. The diffusion gives a decay in the acceleration that depends on the gas's interaction with the surfaces of the \gls{tm} and \gls{eh}.

During the ordinary (as opposed to cold) runs examined in \cite{GlitchSource}, the rate of these glitches was approximately 0.96/day for two test masses. For \gls{lisa}'s three spacecraft, the chance of overlapping glitches is 6.7\% given the 2000 second segments used here. While this probability is not negligible, we consider only isolated glitches here; if this model were incorporated into the global fit, it would automatically account for overlapping glitches.

\section{\label{sec:tdi}TDI Response}
Due to the nature of the \gls{lisa} orbits, the laser links between spacecraft will have unequal and time-varying length. The laser intensity is also significantly diminished over the 2.5 million km mean arm length. Together, these obstacles make typical Michelson interferometry, in which end mirrors are fixed, impossible. Instead, we must use the \gls{tdi} postprocessing technique to measure gravitational waves using \gls{lisa}.

\Gls{tdi} is based on delaying and combining signals from the separate spacecraft as they are passed from one to another via the laser links \cite{tdi}. Its aim is to suppress the laser noise below the level of \gls{gw} signals. For a disturbance to a single \gls{tm} given by velocity timeseries $v(t)$, we can calculate its effect on a constant-, equal-arm length detector as
\begin{eqnarray}
\left.X\right|_{TM_{12}}(t) &=& \frac{f_0}{c}\left[v(t) - 2v(t-4L) + v(t-8L)\right]\\
\left.Y\right|_{TM_{12}}(t) &=& \frac{2f_0}{c}\left[-v(t-7L) + v(t-5L)\right.\nonumber\\
     & & \left. + v(t-3L) - v(t-L)\right]\\
\left.Z\right|_{TM_{12}}(t) &=& 0,
\end{eqnarray}
where $f_0$ is the laser's central frequency, $c$ is the speed of light, $L$ is the arm length in seconds, and $X,Y,Z$ are the 2nd generation Michelson \gls{tdi} channels assuming a glitch on TM$_{12}$, which corresponds to the \gls{tm} on spacecraft 1 pointing toward spacecraft 2. For time-varying, unequal arms, we introduce the operator $d_{ij}$, which delays the signal by the travel time between spacecraft $i$ and $j$ at the time input to the operator, e.g. $d_{23} d_{12} v(t) = v(t - L_{12}(t) - L_{23}(t - L_{12}(t)))$. The response can now be written
\begin{eqnarray}
    \left.X\right|_{TM_{12}}(t) &=& \frac{f_0}{c}\left[d_{21} d_{12} v(t)  \right. \nonumber\\
& & +d_{21} d_{12} d_{31} d_{13} d_{31} d_{13} d_{21} d_{12} v(t) \nonumber\\
& & + d_{31} d_{13} d_{31} d_{13} d_{21} d_{12} v(t) \nonumber\\
& & -d_{21} d_{12} d_{21} d_{12} d_{31} d_{13} v(t) - d_{21} d_{12} d_{31} d_{13} v(t) \nonumber\\
& & \left. -d_{21} d_{12} d_{31} d_{13} v(t) - d_{31} d_{13} v(t) +v(t)\right]\label{eqn:tdix}\\
    \left.Y\right|_{TM_{12}}(t) &=& \frac{2f_0}{c}\left[-d_{21} d_{32} d_{23} d_{32} d_{23} d_{12} d_{21} v(t)\right.\nonumber\\
& & -d_{21} v(t) + d_{21} d_{12} d_{21} d_{32} d_{23} v(t)\nonumber\\
& & \left. +d_{21} d_{32} d_{23} v(t)\right]\label{eqn:tdiy}\\
    \left.Z\right|_{TM_{12}}(t) &=& 0.
\end{eqnarray}
%\begin{eqnarray}
%    \left.X\right|_{TM_{12}}(t) &=& \frac{f_0}{c}\left[d_{13} d_{31} v(t)\right. \nonumber\\
%         & & + d_{13} d_{31} d_{12} d_{21} v(t) \nonumber\\
%         & & \left. - d_{12} d_{21} d_{13} d_{31} d_{13} d_{31} v(t) - v(t)\right]\label{eqn:tdix}\\
%    \left.Y\right|_{TM_{12}}(t) &=& \frac{2f_0}{c}\left[d_{21} v(t) - d_{23} d_{32} d_{21} v(t) \right. \nonumber\\
%         & & - d_{23} d_{32} d_{21} d_{12} d_{21} v(t) \nonumber\\
%         & & + \left. d_{21} d_{12} d_{23} d_{32} d_{23} d_{32} d_{21} v(t)\right]\label{eqn:tdiy}\\
%    \left.Z\right|_{TM_{12}}(t) &=& 0.
%\end{eqnarray}

For glitches on the five other test masses, we can apply simple rotations in the Michelson \gls{tdi} variables to get the response waveform. The rotations amount to permutations and inversions of the $X, Y, Z$ variables given above. They are summarized in Table \ref{tab:xform}.

\begin{table}[h]
    \centering
    \caption{After calculating the response to a glitch with given parameters on TM$_{12}$, the \gls{tdi} data can be transformed for a different injection location using the following relations. Unlisted parameters correspond to $\left.Z\right|_{TM_{12}} = 0$.}
    \begin{tabular}{l r}
        $\left.Y\right|_{TM_{23}}=\left.X\right|_{TM_{12}}$ & 
        $\left.Z\right|_{TM_{23}}=\left.Y\right|_{TM_{12}}$\\
        $\left.Z\right|_{TM_{31}}=\left.X\right|_{TM_{12}}$ & 
        $\left.X\right|_{TM_{31}}=\left.Y\right|_{TM_{12}}$\\
        $\left.X\right|_{TM_{13}}=-\left.X\right|_{TM_{12}}$ & 
        $\left.Z\right|_{TM_{13}}=-\left.Y\right|_{TM_{12}}$\\
        $\left.Y\right|_{TM_{32}}=-\left.Y\right|_{TM_{12}}$ & 
        $\left.Z\right|_{TM_{32}}=-\left.X\right|_{TM_{12}}$\\
        $\left.X\right|_{TM_{21}}=-\left.Y\right|_{TM_{12}}$ & 
        $\left.Y\right|_{TM_{21}}=-\left.X\right|_{TM_{12}}$\\
    \end{tabular}\\
    \label{tab:xform}
\end{table}

After performing these transformations for glitch location, we convert the $(X,Y,Z)$ \gls{tdi} variables to the orthogonal $(A,E,T)$ set \cite{tdi}. These have the advantage of the $T$ channel being insensitive to \gls{gw} signals, but still showing a response to glitches.

\section{\label{sec:likelihood}Maximum Likelihood Estimation}
The \gls{lisa} data will contain numerous overlapping signals with various spectral characteristics \cite{dataanalysis}. However, the glitches examined here are transient effects lasting 10s to 100s of seconds, so we restrict ourselves to short segments of 2000s surrounding the glitch. For the full \gls{lisa} analysis we will need to include the glitch model in the global fit. In aid of that, our search uses a maximum likelihood test, which is evaluated by a \gls{mcmc} routine, just like the global fit.
%To avoid including glitches in a global fit of the \gls{lisa} data, we can detect and remove them before the signal search by using a maximum likelihood test. 
If we suppose that the noise follows a Gaussian distribution in the absence of a glitch, then in the presence of a glitch $G$ with parameters $\beta$, $\Delta v$, and $t_{inj}$ the log-likelihood will be
\begin{eqnarray}
    L &=& \sum_i - \frac{1}{2}\left(\frac{D(t_i)-G(\beta,\Delta v, t_i-t_{inj})-\mu}{\sigma}\right)^2\nonumber\\
        & &- \ln(\sigma\sqrt{2\pi}),
\end{eqnarray}
with $\mu$ and $\sigma$ the mean and standard deviation of $D-G$. In practice, we compute the log-likelihood in frequency space, using the Whittle likelihood as an approximation \cite{Whittle}:
\begin{equation}
    L = -2\sum_i \Re\left(\int_\omega \frac{|F(D_i-G_i)|^2}{S_f^i}\right),
\end{equation}
where $F$ is the Fourier transform, $S_f$ is the PSD of the noise, and $i$ indexes the \gls{tdi} channels.

We run a \gls{mcmc} routine using this likelihood to estimate the maximum likelihood parameters, along with uncertainty values. 
%The application of the \gls{tdi} transform to the glitch waveform results in correlations between the three glitch parameters. Non-orthogonality of parameters can lead to poor \gls{mcmc} performance \cite{mccor}. To reduce these correlations, we replace $\Delta v$ with $\zeta \equiv \Delta v/\sqrt{\beta}$. 
To constrain the walkers, we use flat priors for the ranges $\log\beta\in (-1.6,4)$, $\log\Delta v\in (-22,-7.5)$, and $t_{inj}\in (0,2000)$. We initialize the walkers with a random distribution in full range for $\log\beta$ and $\log\Delta v$, and $t_{inj}\in (475,525)$. We use 400 walkers, which sample the likelihood function for 800 steps, then we analyze the resulting trajectories to estimate the optimum parameters for a glitch in the given data. We cut the initial steps from each walker as burn-in, allowing them to settle in a parameter region. This process results in distributions approximating the true glitch parameters.

\section{\label{sec:gendata}Generated Data}
To test the described method, we generated \gls{lisa} \gls{tdi} data using the Python packages LISA Instrument \cite{LISAIns}, and PyTDI \cite{pytdi}. Glitch parameters were equally distributed in the ranges $\log_{10}\beta\in[-1,1], [1.3,2]$ and $\log_{10}\Delta v\in[-10.5,-8]$, chosen as similar in range to the glitches used in the \gls{ldc} \cite{ldc}, but tuned to test the sensitivity of the described method. We use five values for each of the $\beta$ ranges and the $\Delta v$ range, equally separated in log-space, and take all combinations, resulting in 50 total cases. Each case consists of a 2000 second timespan, sampled at 4 Hz. The glitch is injected at $t = 500$ seconds relative to the start of the segment, but at a random point in the orbits, to allow variation in arm lengths.

A danger in subtracting instrumental transients from data is the possibility of confusing a real signal with a particular waveform. The sine-Gaussian burst waveform gives a \gls{tdi} response that can resemble that of a \gls{tm} glitch \cite{BurstGlitch}, so we also test this detection method with hypothetical burst signals. We consider two burst cases, with parameters chosen to best match a glitch signal, e.g., a source direction resulting in only two of the Michelson \gls{tdi} channels registering it at a significant level.

% For plotting, data are put through a low-pass Kalman filter, to better show the glitch waveform. Neither operation contributes significantly to computational cost.

The search used here allows only time segments with up to one glitch, but if this model were incorporated into the global fit it would account for glitches overlapping with signals and each other. 

\section{\label{sec:results}Results}
\begin{figure}
    \centering
    \includegraphics[width=\linewidth]{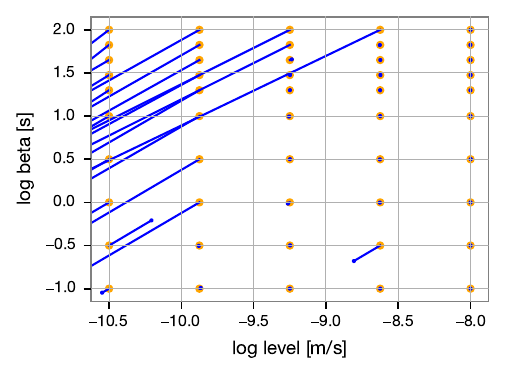}
    \caption{Deviation of recovered parameters (blue) from injected (orange). The points in the upper left corner show systematic large errors due to degeneracy in the parametrization. These points represent low-SNR glitches.}
    \label{fig:params}
\end{figure}

The majority of glitches are recovered with small offsets in their parameters, typically within the range suggested by the distribution of the walkers. A comparison of the injected and recovered parameters is shown in Figure \ref{fig:params}. The largest errors fall in two categories: statistical errors for high $\beta$ with low $\Delta v$, and systematic errors for low $\beta$. For the first case, both high $\beta$ and low $\Delta v$ are associated with lower SNR after \gls{tdi} processing. Combining the two leads to the lowest SNR glitches in the sets considered. %In the second case, the non-orthogonality of these parameters previously discussed comes into play. 
In the second case, as $\beta$ approches and passes the 4 Hz sampling rate, the \gls{tdi} response will fail to capture a change in the shape of the signal, only affecting the overall scale. However, $\Delta v$ changes also result in a change to the response's scale, creating a degeneracy of parameter choices.

\begin{figure}
    \centering
    \includegraphics[width=\linewidth]{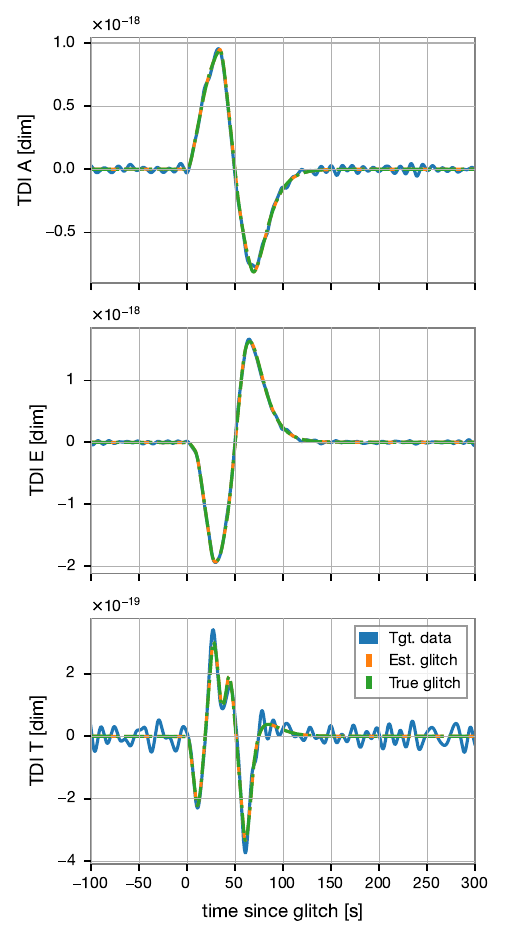}
    \caption{Example of a well-subtracted glitch in the time-domain.}
    \label{fig:goodglitch1_ts}
\end{figure}

\begin{figure}
    \centering
    \includegraphics[width=\linewidth]{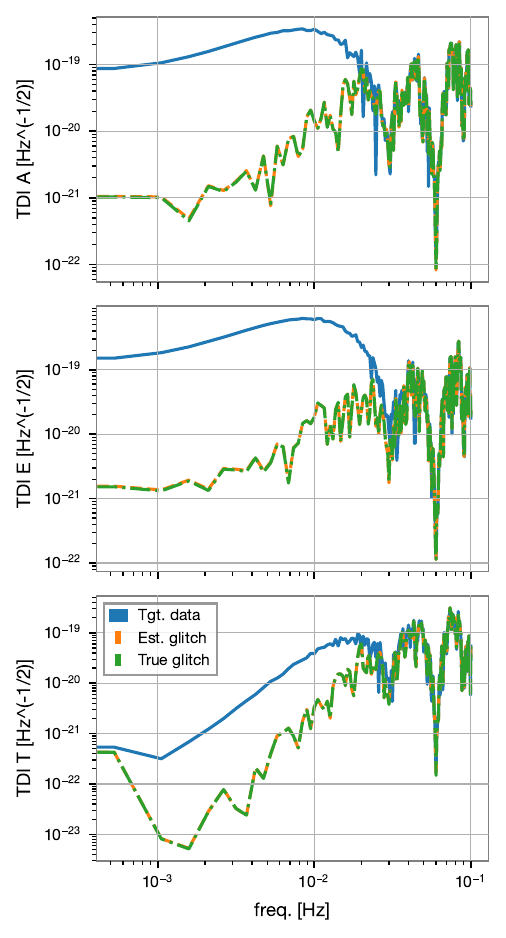}
    \caption{Example of a well-subtracted glitch in the frequency domain. The dashed/dotted curves show the residual spectra after subtracting the given glitch template.}
    \label{fig:goodglitch1_spec}
\end{figure}

\begin{figure}
    \centering
    \includegraphics[width=\linewidth]{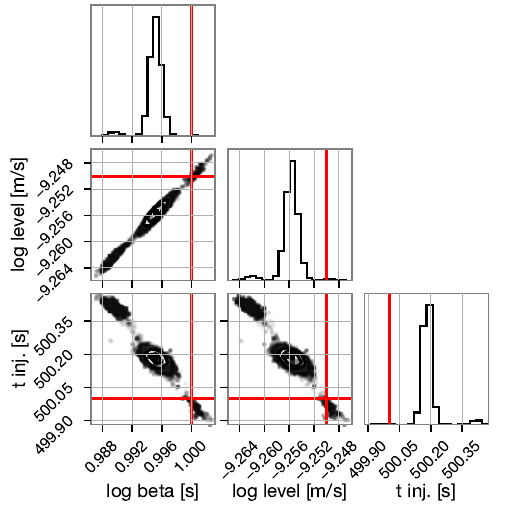}
    \caption{Correlations between parameters for one of the glitches targeted (multimodal instance). The 4 Hz resolution of the timeseries used results in isolated regions of higher likelihood.}
    \label{fig:bincorner}
\end{figure}

An example of a well-subtracted glitch is shown in Figures \ref{fig:goodglitch1_ts} and \ref{fig:goodglitch1_spec}. Some glitches show a multimodal behavior in the walker positions, but this is also attributable to the 4 Hz sampling frequency. We can see how the injection time parameter is separated into these 0.25 second bins, affecting the other parameters through correlation for the case in Figure \ref{fig:bincorner}. %This plot also shows that for cases like this, we still see some non-orthogonality, even using the $\zeta$ parameter in place of $\Delta v$. 

\begin{figure}
    \centering
    \includegraphics[width=\linewidth]{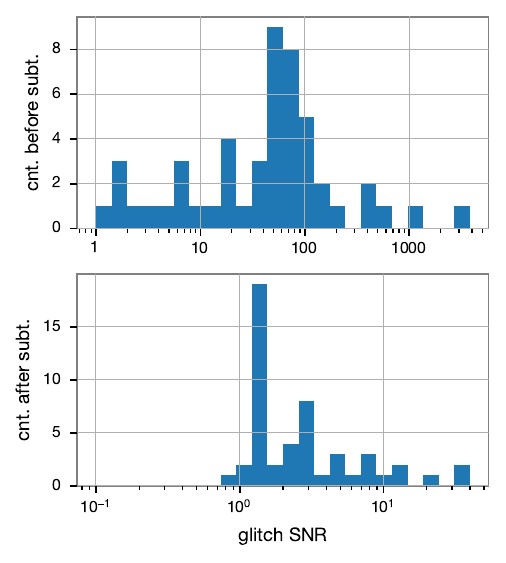}
    \caption{Distribution of maximum SNR of glitch waveforms before and after subtracting maximum likelihood template. After subtraction, the majority of glitches are reduced below SNR 2.}
    \label{fig:snr}
\end{figure}

We calculate a measure of the glitch SNR before subtraction as
\begin{equation}
%    SNR = 2\sqrt{\sum_f\frac{S_D}{S_{D-G}}} = \sqrt{-2L(D)},
    SNR = \sqrt{-2L(D)},
\end{equation}
and after subtraction
\begin{equation}
%    SNR = 2\sqrt{\sum_f\frac{S_{D-\Tilde{G}}}{S_{D-G}}} = \sqrt{-2L(D-\Tilde{G})},
    SNR = \sqrt{-2L(D-\Tilde{G})},
\end{equation}
where $D$ is the \gls{tdi} timeseries including the glitch, $\Tilde{G}$ is the estimated glitch, and $L(X)$ is the log likelihood of data $X$ using the method above. Overall the SNR of the glitch signals is reduced to the level of the noise after subtraction (Figure \ref{fig:snr}).


%\begin{figure}
%    \centering
%    \subfloat[]{
%        \includegraphics[width=0.3\linewidth]{GoodGlitch1_ts.pdf}
%        \label{fig:goodglitch1_ts}}\\
%    \subfloat[]{
%        \includegraphics[width=0.3\linewidth]{GoodGlitch1_spec.pdf}
%        \label{fig:goodglitch1_spec}}
%    \caption{Example of a well-subtracted glitch, with $L = 28170$. \todo{Best way to format these figures? (Also need to fix units)}}
%    \label{fig:goodglitch1}
%\end{figure}

\begin{figure}
    \centering
    \includegraphics[width=\linewidth]{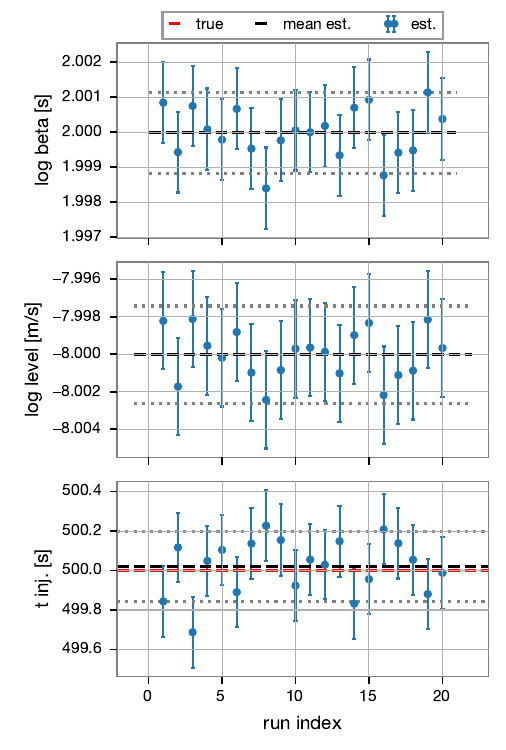}
    \caption{Results of 20 noise realizations for a well-subtracted glitch. Average walker values +/- 1 standard deviations are shown with blue points and errorbars. Red lines show the true parameter values, and black lines show the average values over the blue points. The gray lines are +/- the empirical standard deviation of the blue points (i.e. from measurement of dispersion across realizations). They show excellent agreement with the theoretical standard deviations of the individual runs (blue bars, from \gls{mcmc} sampling).
    %Red lines mark true glitch parameters, black lines show average over all runs, and gray lines show error bars on the average based on the standard deviations of the individual cases.
    }
    \label{fig:meannoise}
\end{figure}

To better understand the discrepancy in glitch parameters for the higher-error cases, we selected several glitches to investigate further. For each glitch, we generated 20 realizations of the noise, and ran the MCMC search on each instance to obtain a distribution of the recovered parameters. Figure \ref{fig:meannoise} shows an example of one of the well-recovered glitches. While the different datasets give varying estimates of the noise, they all overlap with the true parameters of the glitch. In addition to the well-subtracted case shown, we also tested some of the glitches which did not return accurate parameters. By combining the results from multiple noise realizations, we were able to improve the parameter estimation to the same level as the case shown.

%\begin{figure}
%    \centering
%    \includegraphics[width=\linewidth]{CornerNoise.pdf}
%    \caption{Results of 20 noise realizations for a well-subtracted glitch. Lines mark true glitch parameters.}
%    \label{fig:cornoise}
%\end{figure}

For the burst cases, we can use the \gls{tdi} T channel to distinguish the true \gls{gw} signals from spurious glitches. Because \glspl{gw} affect all \glspl{tm} in a correlated manner, the T channel cancels their motion. For a glitch affecting a single \gls{tm} however, we do see a response, e.g. in Figure \ref{fig:goodglitch1_ts}. If we apply our likelihood function to a burst signal before and after subtracting the maximum likelihood glitch, we find the log-likelihood ratio using the T channel favors the burst alone. This suggests our technique can be made robust against mischaracterizing true \gls{gw} signals as glitches.

%\begin{figure}
%    \centering
%    \includegraphics{likhist_trim}
%    \caption{Histogram of log likelihoods for glitch search applied to generated glitches (blue) and bursts (orange). The burst injections show significantly smaller likelihoods than the true glitches. Using tests like these, we can suggest a cutoff value for templates that can be safely subtracted from data.}
%    \label{fig:likhist}
%\end{figure}

\section{\label{sec:concl}Conclusions}
Based on the results from \gls{lpf}, we should expect to see glitches of the type described here in the full \gls{lisa} mission. These glitches have the potential to contaminate \gls{tdi} results in \gls{lisa}'s sensitive band, and we must develop techniques for mitigating their effects. The maximum likelihood method we have described shows promise as a way to find statistically rigorous best-fit parameters for a given glitch, which allow for improved sensitivity after subtraction, often returning the data to its original state. This technique can be incorporated into the \gls{lisa} global fit to allow our existing data pipeline to handle these glitches and reduce their effect on the parameter estimates of true \gls{gw} sources. Because our detection routine is designed for \gls{mcmc}, it should be adaptable as another source type in the global fit. The \gls{ldc} includes datasets with glitches alongside signals, which could be used to test this modified fit. \Gls{lpf} generated valuable information about what we can expect from \gls{lisa}, which lets us prepare for artifacts that could otherwise prevent us from operating \gls{lisa} at its optimal level.
%These results also give insight into how glitches affect the data from \gls{lisa}, and will help in the further development of analysis techniques for the future detector.

\section*{Acknowledgements}
P. W., O. S., W. S. and H. I. were supported by NASA LISA Preparatory Science program, Grant No. 80NSSC19K0324. H. I. acknowledges the Centre Nationale d’Études Spatiales (CNES) for its financial support. Numerical computations were performed on the HiPerGator cluster, UFL. The authors thank the LISA Simulation Working Group and the LISA Artifacts Group for the lively discussions on all simulation related activities.

\bibliography{Biblio}% Produces the bibliography via BibTeX.

\end{document}